\documentclass[manuscript,acmsmall,screen,nonacm]{acmart}

\usepackage{textcomp}
\usepackage{xcolor}
\usepackage{import}
\usepackage{hyperref}
\usepackage{tabularx}
\usepackage{multirow}
\usepackage{algorithm}
\usepackage[noend]{algpseudocode}
\usepackage{amsmath}
\usepackage{graphicx}
\usepackage{subcaption}
\usepackage{mathtools}
\usepackage{booktabs} 

\usepackage{amssymb}
\usepackage{supertabular}
\usepackage{float}
\usepackage{listings}
\usepackage{newfloat}

\usepackage{xcolor} % Optional: 
\usepackage{tikz}

\usepackage{pifont}
\newcommand{\cmark}{\ding{51}} % ✔
\newcommand{\xmark}{\ding{55}} % ✘

\begin{document}

\title{Formal Verification of Secure Encrypted Virtualization}

% \thanks{This work was partially supported by NSF grant SaTC-1936040}
% }

\author{Hansika Weerasena}
\affiliation{%
  \institution{University of Florida}
  \city{Gainesville}
  \state{FL}
  \postcode{32611}
  \country{USA}
}
\email{hansikam.lokukat@ufl.edu}

\author{Amitabh Das}
\affiliation{%
  \institution{AMD}
  \city{Austin}
  \state{TX}
  \postcode{78735}
  \country{USA}
}
\email{amitabh.das@amd.com}

\author{Prabhat Mishra}
\affiliation{%
  \institution{University of Florida}
  \city{Gainesville}
  \state{FL}
  \postcode{32611}
  \country{USA}
}
\email{prabhat@ufl.edu}

\begin{abstract}

Trusted execution environments (TEEs) provide a secure environment for data and code in use, ensuring that they are protected with respect to confidentiality and integrity. Virtual machine (VM)-based TEEs utilize virtualization technology to create isolated execution spaces that can support a complete operating system or specific applications. AMD secure encrypted virtualization (SEV) is a key technology used in confidential computing in the cloud enabling hardware-based memory encryption to protect sensitive data within VMs. However, AMD SEV often operate without formal assurances of their security guarantees. Our research introduces a formal framework for representing and verifying AMD SEV confidential VMs. Specifically, we conduct design-level and property-level abstraction on AMD SEV specification and conduct property checking on the model to ensure confidentiality, integrity and availability. This approach provides a rigorous foundation for defining and verifying key security attributes for safeguarding execution environments. 

\end{abstract}

% \begin{CCSXML}
% <ccs2012>
%    <concept>
%        <concept_id>10002978.10003001.10003003</concept_id>
%        <concept_desc>Security and privacy~Embedded systems security</concept_desc>
%        <concept_significance>300</concept_significance>
%        </concept>
%    <concept>
%        <concept_id>10010147.10010257</concept_id>
%        <concept_desc>Computing methodologies~Machine learning</concept_desc>
%        <concept_significance>500</concept_significance>
%        </concept>
%    <concept>
%        <concept_id>10002978.10003001.10010777.10011702</concept_id>
%        <concept_desc>Security and privacy~Side-channel analysis and countermeasures</concept_desc>
%        <concept_significance>500</concept_significance>
%        </concept>
%  </ccs2012>
% \end{CCSXML}

% \ccsdesc[300]{Security and privacy~Embedded systems security}
% \ccsdesc[500]{Computing methodologies~Machine learning}
% \ccsdesc[500]{Security and privacy~Side-channel analysis and countermeasures}

\keywords{Trusted Execution Environment, Formal Verification, Design Abstraction, Property Checking}

\maketitle

\section{Introduction}
\label{sec:introduction}

Modern cloud computing environments host numerous workloads from mutually untrusted tenants on the same physical infrastructure. This increases the risk of cross-tenant and hypervisor-based attacks, thereby threatening the confidentiality and integrity of data and code executing within virtual machines (VMs). To address these concerns, there is a growing demand for secure hardware primitives that can isolate execution and ensure trust even in the presence of a compromised system software stack. A trusted execution environment (TEE) is a hardware-assisted isolated environment that provides confidentiality and integrity of code and data within a VM, preventing unauthorized access or modification—even by the hypervisor. There are several commercially TEEs, which can be broadly categorized into enclave-based, VM-based, and embedded TEEs. Among thee different classes of TEEs, VM-based TEEs such as Intel TDX, ARM TrustZone, and IBM Z Secure Execution and AMD SEV have emerged as the de-facto standard due to their strong security guarantees, compatibility with legacy applications, and low performance overhead. Unlike enclave-based TEEs (e.g., SGX), VM-based TEEs offer full-VM protection with minimal changes to the application stack, making them suitable for deployment in large-scale public clouds. Embedded TEEs, on the other hand, are typically integrated into mobile and embedded platforms to ensure secure execution at the hardware level. 

While the architectural documentation of VM-based TEEs describes mechanisms for ensuring confidentiality and integrity, it lacks formal guarantees. Given the complexity of modern TEE designs and the potential for subtle implementation flaws, there is a critical need for formal verification of these architectures to validate that they indeed uphold their intended security guarantees. Specifically, we must verify that code and data within a guest VM remain confidential and unmodified throughout the VM’s lifecycle, even in the presence of a malicious hypervisor or system software.

\begin{figure}[htp]
	\centering
        \vspace{-0.1in}
        \includegraphics[width=0.8\columnwidth]{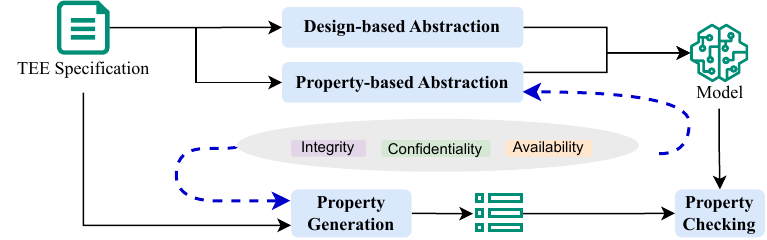}
        \vspace{-0.1in}
        \caption{Overview of formal verification framework.}
	\label{fig:overview_cases}
	    \vspace{-0.1in}
\end{figure}

% Figure~\ref{fig:overview_cases} presents an overview of our formal verification framework for security verification of TEE architectures. We first conduct an abstraction of the TEE architecture that accurately represents the TEE behavior by following the specification. Here, we conduct two types of abstraction; (1). desig-based and (2) property-based. Property-based abstraction phase simplifies the specification, focusing on the essential aspects relevant to confidentiality and integrity and design -based Next, we develop a formal model for VM-based TEE architectures based on the abstraction. Then, we derive properties related to confidentiality and integrity from the TEE specification. Finally, we perform property checking to verify whether the TEE formal model satisfies the specified properties, ensuring that the TEE architecture meets the predefined security criteria on confidentiality and integrity. 

Figure~\ref{fig:overview_cases} presents an overview of our formal verification framework for the security validation of TEE architectures. The process begins with a TEE specification that outlines the expected architecture and behavior of the TEE. From this specification, we apply two complementary forms of abstraction to construct formal models. First, \textit{design-based abstraction} captures all architecturally defined behaviors and transitions described in the specification, independent of any specific property. This abstraction aims to represent the operational semantics of the TEE, forming the baseline for building comprehensive and reusable models. Second, \textit{property-based abstraction} simplifies the system by focusing only on behaviors and components relevant to specific security properties. The resulting formal model is verified using the properties generated from the specification. The property generation phase extracts security properties related to \textit{confidentiality}, \textit{integrity}, and \textit{availability}, each representing a fundamental aspect of system security. Specifically, this paper makes the following major contributions:

\begin{enumerate}
    \item We present a comprehensive formal model that defines the security boundaries of confidential VMs, explicitly considering adversaries with access to advanced attack vectors.
    \item We formally model confidentiality, integrity, and availability notions for VM-based trusted execution environments.
    \item We introduce a detailed formal model for the AMD SEV architecture, developed using the Rosette language.
    \item We critically evaluate the security measures of AMD SEV and compare them with those of other VM-based TEEs, such as Intel TDX.
    \item Finally, we formally verify AMD SEV for confidentiality,  integrity, and availability via multiple properties generated from CIA security notions. 
\end{enumerate}

This paper is organized as follows. Section~\ref{sec:background} provides relevant background and surveys related efforts. Section~\ref{sec:threat} discusses the threat model for VM-based TEEs and defines the formal model for the virtual machine and the adversary. Section~\ref{sec:model-prop} provides formal definitions for confidentiality, integrity, and availability. Section~\ref{sec:sec-analysis} conducts a security analysis of the AMD SEV architecture. Section~\ref{sec:experiments} provides details on the formal modeling of an AMD SEV architecture and discusses the results of the formal analysis. Finally, Section~\ref{sec:conclusion} concludes the paper.

\section{Background and Related Work}
\label{sec:background}

This section provides background for trusted execution environments and AMD SEV. Next, we survey related efforts on verifying trusted execution environments and highlight their limitations.

\subsection{Trusted Execution Environments (TEE)}

TEEs are secure areas within a multiprocessor that provide an isolated execution context to protect sensitive code and data. A TEE ensures that any code running inside it maintains confidentiality and integrity even in the presence of a compromised operating system or privileged software. This isolation guarantees that sensitive operations such as cryptographic key management, attestation, and secure computation are protected from unauthorized access or modification. There are multiple types of TEEs designed with different trust boundaries and deployment models. Enclave-based TEEs, such as Intel SGX, isolate specific regions of an application’s memory, allowing only selected code to access the enclave. These are well-suited for fine-grained protection of small, critical workloads. On the other hand, VM-based TEEs such as AMD SEV and Intel TDX operate at the granularity of full virtual machines, offering broad protection to guest VMs in cloud environments. They secure not only memory but also CPU register states and provide remote attestation, thereby enabling secure multi-tenant computing even in untrusted infrastructure.

\subsubsection{VM-based TEEs}
A virtual machine (VM) abstracts a physical computer, enabling multiple isolated guest operating systems to run on a single host hardware. These VMs are managed by a highly privileged hypervisor that coordinates their execution. To address the threat of a compromised hypervisor, confidential virtual machines (CVMs) have emerged. These are specialized VMs enhanced with hardware-based isolation mechanisms that prevent unauthorized access to VM memory or state—even from privileged software such as the hypervisor or host OS. CVMs are typically realized through a class of TEEs known as VM-based TEEs. Unlike traditional enclave-based TEEs that protect limited code and data regions, VM-based TEEs secure entire virtual machines. They provide complete memory encryption, register state protection, and attestation mechanisms to validate the software stack within the VM. The key architectural difference lies in how the hardware enforces security boundaries.

\subsubsection{AMD Secure Encrypted Virtualization (SEV)}
AMD SEV is a hardware extension introduced by AMD~\cite{amd_sev_api_2020} to enable secure execution of VMs. SEV encrypts the memory of guest VMs using a unique encryption key managed by a secure coprocessor embedded within the AMD processor. This prevents the hypervisor or any external agent from inspecting or modifying the contents of guest memory. The first version of SEV is referred to as \textbf{SEV-base} in rest of the paper. On top of SEV-base, the SEV architecture has evolved through several extensions:
\begin{itemize}
    \item \textbf{SEV-ES (Encrypted State)~\cite{amd_sev_es_2017}:} In addition to memory encryption, SEV-ES ensures that the VM's CPU register state is encrypted during exit event of the VM, thereby preventing leakage of sensitive information through the control flow interface.
    \item \textbf{SEV-SNP (Secure Nested Paging)~\cite{amd_sev_snp_2020}:} SEV-SNP introduces memory integrity verification, page validation, and secure page management through the reverse map table, enabling stronger protections against malicious hypervisor manipulation.
    \item \textbf{SEV-TIO (Trusted I/O)~\cite{amd_sev_tio_2023}:} SEV-TIO extends SEV to protect input/output (I/O) transactions, but it operates outside the scope of our threat model and is not the focus of this work.
\end{itemize}

In this paper, we develop a formal model based on the \textbf{SEV-base} specification and incorporate security-critical features from \textbf{SEV-ES} and \textbf{SEV-SNP}. These include register state protection, page validation, and memory versioning to detect rollback attacks. Our abstraction focuses on guest-visible security behaviors and state transitions throughout the VM lifecycle.

\subsection{Related Work in TEE Verification}

Verification is the process of ensuring that a system behaves as intended and satisfies its specification. Traditionally, security analysis of complex hardware-software systems has relied on simulation-based methods, such as test generation, emulation, or symbolic execution over representative execution paths. While simulation is useful for detecting bugs and testing expected behaviors, it does not guarantee completeness in covering edge cases or unexpected scenarios. In contrast, formal verification offers exhaustive reasoning over all possible system states using mathematical models. This technique allows verification of properties, such as confidentiality, integrity, and correctness across the entire input space. Formal methods typically involve the construction of a formal model followed by property specification and automated checking using tools such as SMT solvers or symbolic evaluators. This section surveys related efforts of TEE verification in three broad categories: static analysis, simulation-based validation, and formal verification.

\subsubsection{Static Analysis}

%Static analysis examines the TEE code without executing the code.  It offers the advantage of early vulnerability detection.  Despite its utility in identifying potential security issues, static analysis lacks a sense of completeness and often generates false positives. 
Google's security review of Intel TDX employed static analysis tools to uncover numerous attack vectors and security issues~\cite{tdxreview}. Security review discovered 81 potential avenues for attacks, confirmed 10 security flaws, and made 5 modifications to enhance the code's defense mechanisms. The review assessed four components of Intel TDX,  including the MCHECK mechanism in BIOS, the non-persistent SEAM loader, the persistent SEAM loader, and the design of the TDX module. 
%Static analysis of the C code was conducted using ``Weggli"~\cite{weggli} and ``Frama-C"~\cite{framac}. ``Weggli" was used to spot design patterns that could potentially compromise the system's security, whereas ``Frama-C" provided a comprehensive framework for static analysis of the codebase. 
However, this approach did not provide formal security guarantees. In fact, this security review highlights the need for formal verification. 

% Our proposed verification framework is complementary to~\cite{tdxreview} since we are using property checking to give a formal guarantee for Intel TDX memory confidentiality and integrity. 

\subsubsection{Simulation-based Validation}
\textcolor{black}{Simulation-based testing methodologies have been used to evaluate the security of TEEs. Google's examination of AMD SEV through simulation-based testing uncovered critical vulnerabilities~\cite{amdreview}. This hands-on approach allows for a practical assessment of TEE security. Simulation-based verification faces the exponential input space complexity to cover all possible scenarios~\cite{witharana2021directed,witharana2022survey}. Nevertheless, the lack of formal security guarantees limits the ability of simulation-based testing to guarantee the security properties of TEE systems.}

\subsubsection{Formal Verification}

Formal methods, such as theorem proving, equivalence checking,  and property checking, have been used for formal verification of TEE architectures~\cite{witharana2024verifying}. For instance, the Moat framework for Intel SGX~\cite{sgx} utilized theorem proving to ensure SGX's confidentiality guarantees~\cite{sinha2015moat}. This abstraction has been extended by~\cite{subramanyan2017formal} for Intel SGX for more properties including integrity, confidentiality, and secure measurement. The authors in~\cite{subramanyan2017formal}  present a verification approach that utilizes a trusted abstract platform (TAP), which formalizes the concept of idealized enclave platforms with a parameterized adversary model.  TEEs based on security enclaves and VMs are different in terms of functionality as well as threat model, and therefore, each of them needs a different verification solution. The TAP framework ~\cite{subramanyan2017formal} provided for verifying security enclaves, which cannot be directly applied for verifying VM-based TEEs. Although both requires modeling of processor and caches, VM-based TEEs rely on many more components and their interactions.

ProveriT~\cite{hu2024proverit} provides a theorem proving solution to formally verify Global Platform TEE common criteria. 
%ProveriT uses Isabelle/HOL theorem prover for the theorems and verification. 
The authors in~\cite{ma2020formal} develop a formal model for memory isolation that includes a detailed formalization of the ARMv8 architecture's hardware components associated with memory isolation, as well as the formalization of a TrustZone monitor that facilitates switching between secure and non-secure worlds. 
A recent study~\cite{sun2020design} introduces a verification methodology for ARM TrustZone using property checking techniques. 
Sardar et al.~\cite{sardar2021demystifying, sardar2023comprehensive} formally specifies the attestation mechanism using ProVerif's specification language. 
This work only focuses on the attestation process, whereas our work focuses on memory confidentiality and integrity of Intel TDX. 
Ozga et al.\cite{ozga2023towards} present a methodology for formally modeling and proving the security of a security monitor in VM-based confidential computing.
Witharana et al.\cite{witharana2024formal} perform property-based formal verification of the Intel TDX trust domain lifecycle.

\subsection{Major Differences in Intel TDX and AMD SEV TEEs}
\label{subsec:tdxvssev}

\textit{Intel TDX} and \textit{AMD SEV} are VM-based trusted execution environments that provide security in the presence of a potentially compromised or untrusted hypervisor. Although they aim to achieve similar goals on confidentiality, integrity, and isolation, their system architecture designs diverge significantly, particularly in how they implement trust, enforcement, and management. We categorize these differences into four key areas:

\vspace{0.1in}
\noindent
\textit{{Centralized firmware-based security monitor vs. hardware-enforced decentralization}:}
A central architectural distinction between Intel TDX and AMD SEV lies in how they enforce trusted execution. TDX introduces a privileged firmware-based \textit{ security monitor} called the TDX Module. This monitor is responsible for enforcing policy on behalf of the Trusted Domain, including managing the VM lifecycle, memory protections, state transitions, and secure interrupt handling. It resides within a highly privileged CPU execution layer, below both the guest and host, and mediates all sensitive operations between the hypervisor and the Trusted Domain. In contrast, AMD SEV does not rely on a centralized software monitor. Instead, it achieves equivalent enforcement through a combination of hardware mechanisms (e.g. memory encryption, register protection, Reverse Map Table) and a trusted coprocessor known as the \textit{AMD security processor (ASP)}. These components collectively perform launch validation, memory ownership tracking, and state isolation without requiring a runtime software monitor. This design leads to a more decentralized enforcement model in which security guarantees are embedded in the hardware datapath itself.

\vspace{0.1in}
\noindent
\textit{{Root of trust and bootloading differences}:} 
The root of trust (RoT) in each architecture also reflects their design philosophies. In Intel TDX, the RoT is centralized and anchored in Intel’s manufacturing infrastructure. Each TDX-enabled processor is provisioned with attestation keys and firmware signed by the Intel Root certificate authority, and the attestation process produces a signed report that can be verified against Intel's global root. This centralized RoT simplifies remote validation but creates a strong dependency on Intel's certificate authority for trust establishment. AMD SEV, on the other hand, takes a decentralized approach. Each platform’s ASP acts as the RoT and is provisioned with unique per-device keys during manufacturing. These keys are used to sign attestation reports locally. While AMD provides a global SEV Root Certificate to verify those reports, the ASP performs attestation independently per system. Consequently, SEV platforms can independently attest to their state without relying on a central authority at runtime, which affects how secure boot flows are rooted and validated.

\vspace{0.1in}
\noindent
\textit{{Attestation process}:}
Attestation in TDX and SEV reflects the trust differences described above. Intel TDX generates an attestation report through the TDX Module and signs it using a global attestation key managed by Intel. In contrast, SEV-SNP’s attestation is performed and signed locally by the ASP using a per-device key provisioned during manufacturing (the Chip Endorsement Key). The TDX attestation key is managed and accessed by the TDX Module but protected in hardware and not embedded in firmware. In AMD SEV, the attestation key is managed and embedded within the ASP. This distinction is important: TDX attestation requires online interaction with Intel’s Attestation Service and relies on a centralized attestation key managed by Intel’s certificate authority. In contrast, SEV-SNP attestation is generated locally and offline by each device’s ASP using unique keys, yet is still verifiable through a certificate chain anchored in AMD’s global CA. This design enables per-device granularity, supports offline validation, and preserves global verifiability.

\vspace{0.1in}
\noindent
\textit{{System-Level architectural distinctions}:} 
A few more key architectural differences arise from these core design choices:
\begin{itemize}
    \item \textit{Lifecycle and memory management}: TDX manages TD lifecycle and memory state through the TDX Module, which tracks ownership and transitions. SEV-SNP relies on hardware-managed metadata like the reverse map table (RMP), and commands issued via application binary interface (ABI) calls to the ASP.
    \item \textit{Hypervisor interaction}: TDX minimizes hypervisor involvement in sensitive operations by funneling them through the TDX Module. In contrast, SEV still allows the hypervisor to issue lifecycle-related commands such as \texttt{ACTIVATE} and \texttt{DEACTIVATE}, although these are verified and controlled by the ASP and hardware access checks.
    \item \textit{Security policy placement}: In TDX, policy logic resides in firmware (TDX Module). In SEV-SNP, policy is enforced structurally through distributed hardware boundaries and validated command sequences.
    \item \textit{Key management}: SEV involves more types keys than TDX, each serving specific roles to enable finer-grained isolation and finer-grained security. For example, SEV-SNP uses a unique per-VM memory encryption key (VEK), generated and managed by the ASP. TDX also uses per-VM memory encryption keys, but these are generated internally by the CPU and bound to each trust domain at runtime. These keys are ephemeral and protected within the CPU. In addition to the VEK, SEV-SNP employs other dedicated keys such as the chip endorsement key (CEK) for attestation and key encryption key (KEK) for key management, showing a complex key hierarchy compared to TDX.
\end{itemize}

\subsection{Limitations of Existing Formal Verification Efforts}

While there are promising formal verification efforts for TEE architectures such as Intel SGX\cite{subramanyan2017formal}, ARM TrustZone~\cite{ma2020formal}, and RISC-V-based TEEs~\cite{ozga2023towards}, existing solutions are not directly applicable to AMD SEV due to their architectural differences. For instance, Intel SGX is an enclave-based TEE that requires developers to explicitly partition and rewrite code to run within enclaves, enabling fine-grained isolation. In contrast, AMD SEV is a VM-based TEE that secures entire virtual machines without requiring code modifications, offering broader compatibility but introducing more complex trust boundaries and dynamic interactions between hardware and software. VM-based TEEs such as AMD SEV and Intel TDX involve intricate state machines, adversarial threat models, and subtler hardware-software trust assumptions. Given the high stakes where a single vulnerability could jeopardize data across many VMs, formal verification becomes critical. It offers the strongest assurance that the architecture correctly enforces isolation and protection guarantees. 

There is a recent effort in formal verification of Intel TDX~\cite{witharana2024formal}. Although Intel TDX and AMD SEV are both VM-based TEEs operating under similar threat models, the mechanisms, microarchitecture, and overall system design they employ to enforce security guarantees differ significantly. For example, they diverge significantly in implementation of security monitor, root of trust, secure boot, attestation, memory tagging mechanisms, page ownership enforcement and key management. 
%These differences are elaborated in Section~\ref{subsec:tdxvssev}, highlighting why the verification approach for Intel TDX~\cite{witharana2024formal} cannot be directly applied for AMD SEV verification. 
These distinctions pose unique challenges when adapting verification strategies across the two architectures and highlight the need for architecture-specific formal modeling and verification, which is the focus of this paper.

\section{Threat Model for VM-based TEE}
\label{sec:threat}

\begin{figure}[htp]
	\centering
 % \vspace{-0.2in}
\includegraphics[width=0.7\columnwidth]{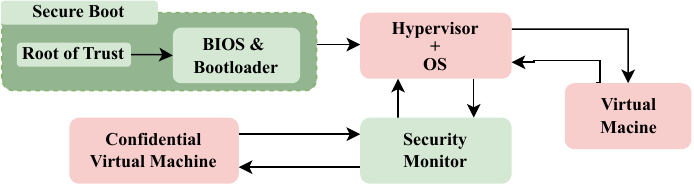}
 \vspace{-0.1in}
	\caption{Overview of threat model for VM-based confidential computing.}
	\label{fig:confidentialt_computing_overview}
	    \vspace{-0.05in}
\end{figure}

Figure~\ref{fig:confidentialt_computing_overview} shows the basic building blocks required for VM-based TEE architecture.  It starts with a secure boot process; the system relies on a foundational security mechanism known as the root of trust (RoT), complemented by the principle of a chain of trust. The RoT is pivotal for ensuring that only authenticated and integrity-verified firmware and software are loaded for execution. It achieves this through the provision of essential cryptographic functions and services. Initially, the RoT verifies the integrity and authenticity of the bootloader and establishes the first link in the chain of trust. Once the bootloader is authenticated, it securely loads and verifies the firmware, setting the stage for the execution environment. Hypervisor can create and manage VMs. In a type~2 hypervisor configuration, a host OS will work alongside the hypervisor, while type~1 will have only a hypervisor. A fundamental element of a VM-based confidential computing framework is the security monitor. It operates at a low level, closely interacting with the hypervisor and host operating system to monitor and control access to resources, manage permissions, and ensure isolation between different confidential VMs. The security monitor's primary objectives include preventing unauthorized access to sensitive data, ensuring that software components cannot interfere with each other maliciously, and enforcing compliance with security protocols. AMD SEV~\cite{amdreview} has a security monitor.

In AMD SEV, there are three main operational flows:

\begin{itemize}
    \item \textit{Platform provisioning flow}: Resets the SEV platform to a clean factory state, initializes the secure processor, and installs the Platform Endorsement Key certificate signed by a trusted Certificate Authority.
    \item \textit{Guest launch flow}: Initializes and encrypts a confidential VM, securely loads its memory pages, measures its initial state, and finalizes its launch with attestation support to establish trust in the VM’s configuration.
    \item \textit{Live migration flow}: Enables securely migrating a running SEV-protected VM between physical platforms, ensuring that memory encryption keys and VM state are protected throughout the migration process.
\end{itemize}

In our threat model, we assume that the platform provisioning flow has already been completed correctly by a trusted cloud provider. Thus, we do not model and verify the provisioning process. We focus on modeling and verifying guest launch and live migration flows, capturing their security-critical operations. This is justified because platform provisioning is a rare, privileged operation performed by the cloud provider during platform setup or reallocation. It is reasonable to assume that this process is executed correctly and securely, allowing it to be excluded from guest-level threat models. The rest of the section introduces formal models for virtual machine and  adversary in the context of trusted execution.

\subsection{Formal Model of Virtual Machines}

An VM operates within a host system, leveraging allocated resources such as CPU cores, memory, and virtualized I/O. Upon initialization, the VM state is captured in a foundational snapshot, preserving its identifier, OS image, virtual address mapping, memory allocation, and relevant code and data pages.

\begin{itemize}
    \item \textbf{VM inputs}: These can be external or internal. External inputs deterministically alter the VM’s execution state, while internal inputs (such as executed code) have variable effects depending on runtime conditions.
    \item \textbf{VM states}: The VM’s execution state, denoted as \(S_{VM}(m)\), represents a subset of the host machine state \(m\). This state includes mappings of virtual addresses, register values, and other internal execution attributes.
    \item \textbf{VM transitions}: The VM undergoes deterministic state transitions, where the next state is a function of the current state and the received input.
    \item \textbf{VM outputs}: The VM generates observable outputs, which may be encrypted in memory or decrypted within the processor before execution.
\end{itemize}

\subsection{Formal Model of the Adversary}

In the threat model, we consider a privileged adversary with elevated system privileges capable of influencing VM execution. The adversary can pause, terminate, and initiate VMs, providing significant control over the system. Furthermore, the adversary can manipulate both the system state \(m\) and their own adversary state \(A_{V}(m)\) by executing arbitrary instructions. However, the security mechanisms ensure that the adversary cannot directly modify the confidential VM state \(S_{V}(m)\), preserving data integrity. Despite these capabilities, the adversary’s ability to observe VM execution is constrained by the security monitor, which enforces isolation and protection mechanisms. The function \(\textit{monitor}_{V}(m)\) defines the extent to which the adversary can observe a VM’s state based on the machine state \(m\). For example, if a VM’s computation results are externally consumed, the adversary’s observed state may be equivalent to the VM’s output (i.e., \(\textit{monitor}_{V}(m) = O_{V}(m)\)).

A VM execution sequence can be represented as an unbounded sequence of machine states:
\[
\sigma = (m_{0}, m_{1}, \dots, m_{n})
\]
where each transition corresponds to a deterministic computation step governed by the VM’s execution model.
%\textcolor{red}{Hypervisor rollback attacks (e.g., reverting to an old VM state).}

\section{Formal Modeling of Properties}
\label{sec:model-prop}

This section introduces the formal models for CIA (Confidentiality, Integrity, Availability) notions in the context of VM-based trusted execution.

\subsection{Formal Model for Confidentiality}
\label{subsec:conf-model}

\begin{figure}[htp]
\vspace{-0.2in}
	\centering
    \includegraphics[width=0.6\columnwidth]{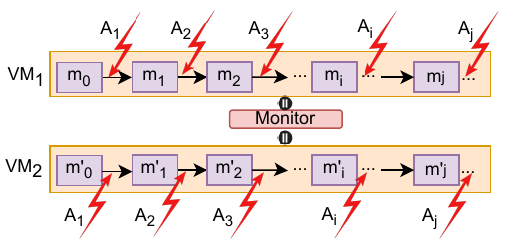}
    \vspace{-0.2in}
	\caption{Overview of the confidentiality property: Two VMs, \(V_1\) and \(V_2\), start from equivalent initial states but diverge due to internal computations. The adversary, constrained by \(\textit{monitor}_{V}(m)\), cannot distinguish between them.}
    \vspace{-0.1in}
	\label{fig:confidentiality}
\end{figure}

As noted, confidentiality threats are handled through the hardware-based memory encryption present in all current SEV technologies. This prevents an untrusted component, such as the hypervisor or a DMA-capable device, from being able to directly read the plaintext inside a VM (except of course in cases where the VM has opted to allow untrusted access to a page). The SEV-ES technology added confidentiality protection for VM register state, encrypting this state when the VM exits back to the hypervisor. This protection exists in SEV-SNP as well. Confidentiality ensures that an adversary cannot distinguish between the execution of two different VMs, \(VM_1\) and \(VM_2\), solely through external monitoring. This property is illustrated in Figure \ref{fig:confidentiality}, where two VMs start from equivalent initial states (\(m_0 = m'_0\)) but follow different computational paths due to distinct internal computations. $A_1, A_2, \dots, A_j$ is a sequence of adversary actions throughout the VMs execution. Despite these variations, the adversary, observing through \(\textit{monitor}_{V}(m)\), cannot differentiate the execution behavior of the two VMs. The confidentiality property can be formalized as:

\begin{align*}
&\forall \sigma_1, \sigma_2. \bigg( A_{v1}(\sigma_1[0]) = A_{v2}(\sigma_2[0]) \land \\
&\forall i. \big(\text{curr}(\sigma_1[i]) = \text{curr}(\sigma_2[i]) \land I_{v1}(\sigma_1[i]) = I_{v2}(\sigma_2[i])\big) \land \\
&\forall i. \big(\text{curr}(\sigma_1[i]) = v\big) \Rightarrow \\
&\text{monitor}_{v1}(\sigma_1[i + 1]) = \text{monitor}_{v2}(\sigma_2[i + 1]) \bigg) \\
&\Rightarrow \bigg( \forall i. A_{v1}(\sigma_1[i]) = A_{v2}(\sigma_2[i]) \bigg)
\end{align*}

\subsubsection{Data Confidentiality}

Data confidentiality is a key property in TEE. Confidentiality should ensure that the sensitive data is accessed only by the authorized parties. Some of the confidentiality properties can be (1) hypervisor should not be able to read the private VM memory, (2) after exiting the VM, VM register state should not be readable, and (3) untrusted devices should not be able to read VM memory~\cite{TechnicalReportAMD}. For example, we can use the property $P_1$ to check the data confidentiality. Different side-channel attacks such as PRIME+PROBE can be used to track data in VM access. Similarly, to protect the system trace bus from getting used as a side-channel access when a secure virtual machine is scheduled, we need to generate a property similar to $P_2$.

\vspace{0.05in}
$P_1:  E(hypervisor \implies \neg read(private\_vm\_memory))$
% \qed
 
\vspace{0.05in}
$P_2:  E(state(secure\_vm) \implies \neg access(system\_trace))$
% \qed

%\vspace{-0.1in}
\subsubsection{Code Confidentiality}
%\vspace{-0.05in}
In addition to protecting data, some TEEs protect code while in use from being viewed by unauthorized entities (e.g., $P_3$).

\vspace{0.05in}
$P3: E(unauthorized\_user \implies \neg code\_access(IP))$
% \qed

\subsection{Formal Model for Integrity}
\label{subsec:int-model}

\begin{figure}[htp]
\vspace{-0.2in}
	\centering
	\includegraphics[width=0.6\columnwidth]{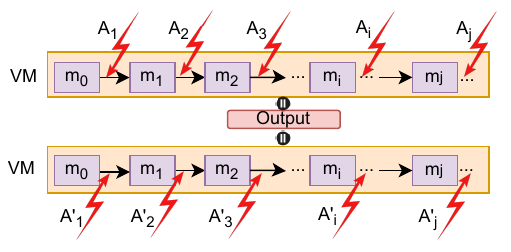}
    \vspace{-0.2in}
	\caption{Overview of the integrity property: The VM’s execution remains unchanged despite different adversary actions (\(A_1\) and \(A_2\)), ensuring consistent output and execution trace.}
    \vspace{-0.1in}
	\label{fig:integrity}
\end{figure}

SEV-SNP technology is designed to protect against integrity attacks, which include data replay, corruption, re-mapping, and aliasing-based attacks. The guarantee that a VM sees the data it last wrote implies that all these attack vectors must be prevented. The integrity property ensures that a VM’s execution remains consistent despite different adversary actions. As shown in Figure~\ref{fig:integrity}, two VM execution sequences start from equivalent initial states (\(m_0 = m'_0\)), but the adversary applies different actions (\(A_1\) and \(A_2\)) during execution. Despite these variations, the VM receives the same sequence of inputs, ensuring that its execution trace and final output remain unaffected. This guarantees that adversarial manipulation does not compromise the correctness or consistency of the VM’s execution. By enforcing confidentiality and integrity, the formal model ensures that VMs can execute securely within an adversarial environment, preserving both data privacy and execution correctness. The integrity property can be formalized as:

\begin{align*}
&\forall \sigma_1, \sigma_2. \bigg( S_{\text{V}}(\sigma_1[0]) = S_{\text{V}}(\sigma_2[0]) \land \\
&\quad \forall i. \big(\text{curr}(\sigma_1[i]) = \text{V}\big) \Leftrightarrow \big(\text{curr}(\sigma_2[i]) = \text{V}\big) \land \\
&\quad \forall i. \big(\text{curr}(\sigma_1[i]) = \text{V}\big) \Rightarrow I_{\text{V}}(\sigma_1[i]) = I_{\text{V}}(\sigma_2[i]) \bigg) \\
&\Rightarrow \bigg( \forall i. S_{\text{V}}(\sigma_1[i]) = S_{\text{V}}(\sigma_2[i]) \land  O_{\text{V}}(\sigma_1[i]) = O_{\text{V}}(\sigma_2[i]) \bigg)
\end{align*}

\subsubsection{Data integrity}
%\vspace{-0.05in}
Data integrity should ensure that the data is consistent  and accurate throughout the process. Some of the integrity checks that can be performed are: (1) Only the owner of a memory page can write that page (e.g., $P_3$), and (2) Every physical memory page can map only to a single guest page at one time~\cite{TechnicalReportAMD}. Different types of attacks that can be preformed to breach the integrity are replay protection, data corruption, memory aliasing, and memory re-mapping. 

% VM Memory 
% Example attack: Hypervisor reads private VM memory 

% VM Register State 
% Example attack: Read VM register state after VMEXIT 

% DMA Protection Example attack: Device attempts to read VM memory 

% Integrity 
% Replay Protection 
% Example attack: Replace VM memory with an old copy 

% Data Corruption 
% Example attack: Replace VM memory with junk data 

% Memory Aliasing 
% Example attack: Map two guest pages to same DRAM page 

% Memory Re-Mapping 
% Example attack: Switch DRAM page mapped to a guest page

\vspace{0.05in}
% $P_2: E(autnorized\_state |=> protected\_state \&\& consistent\_data(autnorized\_state, protected\_state))$
$P_4: E(write(page) \implies user==owner)$
% \qed

%\vspace{-0.1in}
\subsubsection{Code Integrity}
%\vspace{-0.05in}
Code integrity ensures that code in the TEE cannot be replaced or modified by unauthorized entities. Although the data is identified as the most sensitive information on a VM, sometimes code-specific information can be vital, such as the version of the running code at the moment. TCB rollback attacks can be performed to degrade firmware to older vulnerable versions. We need to check a property similar to $P_5$.

\vspace{0.05in}
% $P_4: \neg E(unutnorized\_user \implies modify(IP))$
$P_5: E(version(firmware)==secure\_version )$
% \qed

\color{blue}

% %\vspace{-0.1in}
% \subsubsection{Availability}
% %\vspace{-0.05in}
% Availability ensures that hypervisor retains control of the system and the guest VM is not able to deny the hypervisor from running. Some of the availability checks that can be performed are: (1) Checking the exit and yield for any guest (e.g., $P_6$), and (2) Running different guests on hypervisor. 

% \vspace{0.05in}
% \noindent {\bf Property 6:}
% $P_6: E(user==guest\&\& state(exit) \implies state(success))$
% % \qed

% % \subsubsection{Authenticated Launch}
% % In some TEEs authorization or authentication checks are mandatory before launching a process.

% % \vspace{0.05in}
% % \noindent {\bf Property 5:}
% % $P_5: E(launch(process) |=> authenticated\_user)$
% %  \qed

% %\vspace{-0.1in}
% \subsubsection{Attestability}
% %\vspace{-0.05in}
% TEE ensures secure storage and operations on encryption keys and other related assets. Attestability can be breached using architectural side-channel attacks on CPU data structures and fingerprinting attacks. Therefor, we need to check  $P_7$.

% \vspace{0.05in}
% \noindent {\bf Property 7:}
% $P_7: E(signature(IP, public\_key)==valid)$
% % \qed

\color{black}

\subsection{Formal Model for Availability}
\label{subsesc:avil-model}

% There are two aspects of availability with any virtualization platform. The first is ensuring that the hypervisor retains control of the system, and the guest VM is not able to deny the hypervisor from running or otherwise render the physical machine unusable. 

% First, its important to note that availability is not primary focus of TEEs but in actual use cases we cannot simply forget about availability. For example, we do not want our banking application to indefinitely halt. Formalizing availability for VM-based TEE requires availability in two angles: host-side safety and guest-side liveness.

While availability is not the primary design focus of TEEs, it is a critical requirement in practical deployments. For instance, applications such as online banking or healthcare services must not be allowed to stall indefinitely or they do not want to affect each others availability by halting the hypervisor. To formally capture availability in the context of VM-based TEEs, we consider two complementary dimensions: \textit{host-side liveness} and \textit{guest-side liveness}. \textit{Host-side liveness} ensures that a guest VM cannot indefinitely prevent the host from reclaiming resources. Conversely, \textit{guest-side liveness} ensures that the hypervisor or platform cannot indefinitely suppress or delay progress of a guest's execution, thereby guaranteeing that legitimate workloads are not starved of CPU resources or delayed.  
We define a unified model that supports both host-side safety and guest-side liveness in the presence of a privileged adversary. Let \( curr(m) \in \{v, h\} \) denote whether the VM (\( v \)) or hypervisor (\( h \)) is executing at state \( m \).  
Let \( Sv(m) \) be the VM lifecycle state (e.g., \texttt{RUNNING}, \texttt{DEACTIVATED}), and \( Ready(Sv(m)) \) indicate that the guest is prepared to execute. Availability property can be formalized as:

\[
\forall i.\;
\left\{
\begin{aligned}
curr(m_i) = v &\Rightarrow \exists j \ge i.\; curr(m_j) = h \\
curr(m_i) = h \land Ready(Sv(m_i)) &\Rightarrow \exists j > i.\; curr(m_j) = v \land Sv(m_j) \ne Sv(m_i)
\end{aligned}
\right.
\]

\begin{figure}[htp]
	\centering
    \vspace{-0.2in}
	\includegraphics[width=0.6\columnwidth]{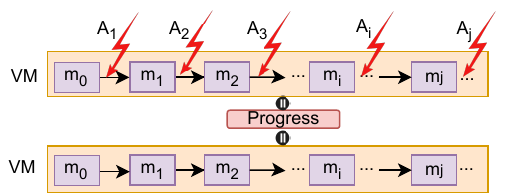}
    \vspace{-0.1in}
	\caption{Overview of the availability property: The VM's execution eventually reaches the same state despite adversary actions (\(A_1, A_2, \ldots, A_j\)), ensuring that guest progress is not indefinitely denied and termination remains achievable under interference.}
    \vspace{-0.1in}
	\label{fig:availability}
\end{figure}

This ensures that the hypervisor can regain control all the time, regardless of guest behavior, and ready guest must eventually be scheduled and make progress. However, it is important to note that \textit{guest-side liveness} cannot be guaranteed in any VM-based TEE with an untrusted hypervisor, as the hypervisor retains control over CPU scheduling and can arbitrarily delay or suppress guest execution. Figure~\ref{fig:availability} models \textit{guest-side liveness} availability as a two-trace guarantee. VM \( VM_1 \) proceeds to state \( m_j \) under normal conditions, while \( VM_2 \) is subject to adversarial actions. If \( VM_1 \) terminates, then \( VM_2 \) must eventually reach an equivalent state \( m_j \), ensuring guest progress is not indefinitely denied.

% \subsection{Formal Model for Attestability}

\section{Security analysis of AMD SEV and its extensions}
\label{sec:sec-analysis}

AMD SEV is designed to provide memory encryption and integrity protection for virtualized workloads, ensuring isolation between VMs and the hypervisor. The security of SEV relies on multiple components of AMD secure processor (ASP), including platform management, guest lifecycle operations, memory encryption, attestation and key management. This section analyzes the security mechanisms in SEV-base and evaluates how these components collectively ensure the confidentiality and integrity of VMs. Finally, we discuss AMD SEV-ES and AMD SEV-SNP, two extensions to the base SEV architecture that introduce enhanced security features and stronger isolation guarantees.

\subsection{AMD Secure Processor (ASP)}

The ASP is a dedicated security co-processor integrated into AMD CPUs that serves as the root of trust for AMD SEV. It operates independently of the main processor cores and is responsible for managing cryptographic operations critical to platform and guest security. The ASP runs SEV firmware that handles key management, platform identity provisioning, attestation, and memory encryption setup. During the platform initialization phase, the ASP processes commands such as \texttt{INIT}, \texttt{PEK\_CSR}, and \texttt{PDH\_CERT\_EXPORT} to establish a trusted platform state. When launching or migrating SEV-protected guests, it handles commands like \texttt{LAUNCH\_START}, \texttt{ACTIVATE}, and \texttt{LAUNCH\_UPDATE\_DATA}, ensuring that guest memory is encrypted and isolated from higher-privileged software such as the hypervisor. By isolating sensitive cryptographic functions and keys from the main system software, the ASP enables strong confidentiality guarantees even in the presence of a compromised hypervisor. Note that the actual guest VM will run on specific AMD EPYC server processors with SEV extensions. 

\subsection{SEV Platform Security and Integrity Verification}

The SEV platform refers to the physical host system that includes the AMD secure processor and SEV firmware responsible for managing the security features of virtualization. The SEV platform acts as the root of trust and handles cryptographic operations such as key generation, memory encryption, and attestation support. While guests are transient and tenant-specific, the SEV platform is persistent and maintained by the cloud provider. It provides foundational services that enable guests to operate securely in a potentially untrusted environments. SEV includes mechanisms for platform security, ensuring that the underlying hardware is trustworthy and resistant to unauthorized modifications. This is managed through several key operations:

\begin{itemize}
    \item \textit{Platform lifecycle management:} SEV supports commands like \texttt{INIT} and \texttt{SHUTDOWN} to manage platform state, ensuring consistent security from boot to shutdown.
    \item The platform uses the \textit{platform endorsement key (PEK)} and \textit{platform Diffie-Hellman (PDH)} key to establish cryptographic identity, authenticate to external parties, and set up session keys during guest launch.
    \item \textit{Device authenticity}: Each SEV-enabled processor contains a \textit{chip endorsement key (CEK)}, which is signed by AMD and used to prove that the hardware is genuine. This prevents attackers from emulating SEV platforms using counterfeit or tampered hardware.
\end{itemize}

The use of these cryptographic mechanisms ensures that the platform operates within a trusted execution environment, preventing adversaries from tampering with system initialization and memory encryption settings.

\subsection{Guest Lifecycle Security}

A critical component of SEV security is how guest VMs are securely launched, managed, and decommissioned. 
%Figure~\ref{fig:life_cycle} illustrates the first few stages of SEV guest lifecycle, which securely provisions and starts a virtual machine (VM) with memory encryption and remote attestation support. 
The flow spans multiple parties: \textit{Guest Owner, Cloud Provider, Hypervisor, and the AMD Secure Processor}. The goal is to ensure that the guest VM is launched in a trusted state, verified by the guest owner, with all memory encrypted and protected from the hypervisor. First, the guest owner, cloud provider, and hypervisor come up with agreement on boot image and its cryptographic hash is computed to enable attestation later. The key security properties ensured during different lifecycle phases include:

% \begin{figure*}[tp]
%     \centering
%     \includegraphics[width=0.90\textwidth]{./figures/life_cycle.pdf}
%     \caption{First few steps of launching guest flow.}
%     \label{fig:life_cycle}
% \end{figure*}

\begin{itemize}
    \item \textit{Secure VM launch:} SEV provides commands to initialize a VM with encrypted memory and control its launch process. This ensures that memory is encrypted before guest execution begins, and secrets are only injected after the guest is securely set up.
    
    \item \textit{Activation and address space isolation:} SEV assigns an address space identifier (ASID) to each VM and associates it with a unique memory encryption key. This ensures strong memory isolation and prevents unauthorized access from other guests or the hypervisor.
    
    \item \textit{Attestation:} SEV supports cryptographic attestation of the guest's launch state through a measurement process. A remote guest owner can verify the integrity of the loaded guest image to ensure that it matches the expected configuration before injecting secrets.
    
    \item \textit{Secure migration and snapshotting:} When migrating or snapshotting a VM, SEV ensures that memory contents remain encrypted and protected from the hypervisor or network adversaries during transit and storage.
    
    \item \textit{Guest decommissioning:} Upon VM termination, SEV invalidates the guest’s encryption keys to ensure that memory contents cannot be recovered or decrypted after the guest is shut down or reallocated.
    
    \item \textit{Secure debugging:} SEV allows debugging only when explicitly permitted by the guest owner’s policy. This prevents unauthorized memory access or inspection by privileged software during runtime.
\end{itemize}

These mechanisms collectively secure the confidential VM lifecycle, defending against rollback and cloning attacks, enforcing isolation throughout execution, and preventing data leakage during live migration and decommissioning.

% \subsection{Memory Encryption and Isolation}

% A fundamental aspect of SEV security is its ability to encrypt VM memory with a unique encryption key, preventing unauthorized access by the hypervisor or other VMs. This is enforced by the SEV firmware running on the AMD Secure Processor (ASP), which manages key provisioning and data encryption transparently. Each VM is assigned a distinct \textit{VM Encryption Key (VEK)}, ensuring that no two VMs can decrypt each other's memory, thereby mitigating cross-VM attacks.

% Memory encryption ensures that even if an attacker gains privileged access to the hypervisor, they cannot extract plaintext data from the VM’s memory. This design strengthens security against cold boot attacks, memory scraping, and side-channel attacks by ensuring that memory remains encrypted in DRAM.

\subsection{Memory Encryption and Isolation}

A core security feature of SEV is its ability to transparently encrypt the memory of VMs, thereby protecting against unauthorized access by the hypervisor or other software components. The encryption itself is performed by the memory controller using a unique \textit{VM encryption key (VEK)} assigned to each VM. The SEV firmware running on the ASP is responsible for provisioning these keys, enforcing access control, and managing guest lifecycle commands. The ASP ensures that the VEK is securely generated and loaded into hardware registers, after which the memory controller uses it to encrypt and decrypt memory lines as they leave or enter the CPU. 

This design ensures strong isolation between co-resident VMs and the hypervisor. Because memory encryption is tied to per-VM keys, no VM can decrypt or observe the memory contents of another, and the hypervisor cannot access guest memory in plaintext—even with full system privileges. This mitigates a wide range of attacks, including cold boot attacks, memory scraping, and certain types of side-channel attacks that rely on access to unencrypted memory. By ensuring that all memory leaving the CPU is encrypted, SEV strengthens the security boundary around each guest, enabling safe execution even in environments where the hypervisor is untrusted. This memory protection is integral to SEV's threat model and serves as the foundation for higher-level guarantees such as secure launch, migration, and decommissioning.

\subsection{Key Management and Cryptographic Root of Trust}

The security of SEV is grounded in a hierarchical key management architecture that anchors trust to AMD’s hardware root of trust. This system enables the platform to securely provision, verify, and isolate cryptographic operations across both platform and guest contexts. SEV utilizes a diverse set of cryptographic keys, each serving a distinct role in securing different parts of the system:

\begin{enumerate}
    \item \textit{Platform diffie-hellman key (PDH):} Used to establish secure channels between the SEV firmware and remote entities such as guest owners.
    
    \item \textit{Platform endorsement key (PEK):} Authenticates the PDH and cryptographically binds it to the platform’s identity.
    
    \item \textit{Chip endorsement key (CEK):} A unique per-processor key signed by AMD, ensuring the authenticity of SEV-enabled hardware.
    
    \item \textit{AMD signing key (ASK) and AMD root key (ARK):} Root signing keys managed by AMD that certify the CEK and form the foundation of SEV's trust hierarchy.
    
    \item \textit{Owner certificate authority (OCA) Key:} Used by cloud providers or platform owners to endorse PEKs and assert platform ownership in custom deployments.
    
    \item \textit{Transport integrity key (TIK) and transport encryption key (TEK):} Session-specific keys that ensure the confidentiality and integrity of memory pages during operations like migration and secret injection.
    
    \item \textit{Key encryption key (KEK) and key integrity key (KIK):} Wrap the TEK and TIK for secure transport during session establishment.
    
    \item \textit{VM encryption key (VEK):} A unique, ephemeral key provisioned per VM, used by the memory controller to encrypt and decrypt VM memory in hardware.
\end{enumerate}

This hierarchical design ensures that all security-critical operations—such as attestation, key injection, and memory protection—can be cryptographically validated. As each key is bound to a higher-trust authority, an attacker would need to break the entire chain of trust.

\subsection{AMD SEV Encrypted State (SEV-ES)}

AMD SEV encrypted state (SEV-ES) is an architectural extension to the base SEV model that enhances guest protection by encrypting and integrity-protecting the CPU register state during VM context switches. While base SEV ensures memory confidentiality through per-VM encryption, it does not protect the CPU registers, which remain exposed to the hypervisor during VM exits. SEV-ES addresses this limitation by ensuring that all general-purpose and control registers are encrypted and inaccessible to the hypervisor, thereby shrinking the guest attack surface.

SEV-ES strengthens the confidentiality and integrity of a guest VM by preventing register-based attacks. A malicious hypervisor can exploit unprotected registers to read sensitive data (e.g., cryptographic keys in XMM registers), inject malicious control flow (e.g., by modifying the instruction pointer), or replay previous register states to corrupt execution. By encrypting and verifying register contents on every VM exit and re-entry, SEV-ES mitigates these threats. It prevents silent data exfiltration, unauthorized state manipulation, and replay attacks that could compromise control flow integrity. SEV-ES extends the confidentiality and integrity guarantees of AMD-base as follows:  
\begin{itemize}
    \item \textit{Confidentiality:} Protects sensitive register contents from hypervisor inspection.
    \item \textit{Integrity:} Detects unauthorized modifications or replay of register state by the hypervisor.
    % \item \textit{Availability:} While SEV-ES does not directly improve availability, it prevents control flow corruption that could cause unintended halts or misbehavior.
\end{itemize}

By protecting the register state alongside encrypted memory, SEV-ES significantly reduces the guest VM's exposure to a compromised or malicious hypervisor and moves closer toward full VM state isolation.

\subsection{AMD SEV Secure Nested Paging (SEV-SNP)}

AMD secure nested paging (SEV-SNP) extends SEV-base and SEV-ES by addressing their remaining trust dependency on the hypervisor, specifically the assumption that the hypervisor maintains correct and secure nested page tables. SEV-SNP introduces hardware-enforced memory integrity mechanisms that ensure isolation, even in the presence of a fully compromised hypervisor. At the core of SEV-SNP is the \textit{reverse map table (RMP)}, a data structure managed by the ASP that records ownership, assignment status, and validation state for each physical memory page. This enables the platform to track and verify how memory pages are assigned across guest VMs, the hypervisor, and firmware components. Each memory access or mapping operation is checked against this metadata to prevent unauthorized remapping, aliasing, or injection attacks.

\begin{figure}
%\vspace{-0.35in}
\centering
\includegraphics[width=3.0in]{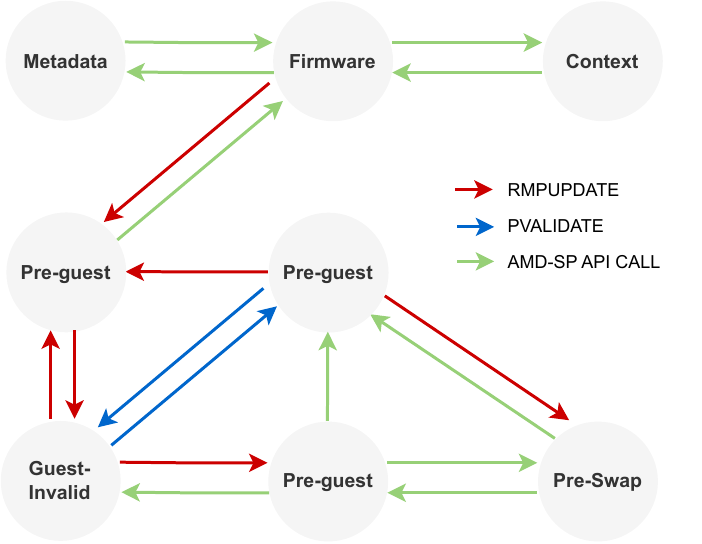}
\vspace{-0.2in}
\caption{SEV-SNP page state transitions~\cite{TechnicalReportAMD}.}
\label{fig:firmwarefixed}
\vspace{-0.2in}
\end{figure}

SEV-SNP enforces memory access policies through a finite state machine that governs transitions between page states such as Guest-Valid, Pre-Guest, Guest-Invalid, Firmware, and others. Figure~\ref{fig:firmwarefixed} illustrates the SEV-SNP page state transitions. Each transition is mediated through a specific mechanism: 
\texttt{RMPUPDATE} (red arrows) is used to update the RMP, \texttt{PVALIDATE} (blue arrows) ensures that memory pages have been validated, and AMD-SP firmware APIs (green arrows) enforce additional platform policies. For example, the hypervisor must explicitly validate pages before transitioning them into a guest-accessible state. Invalid or unauthorized transitions are rejected by the hardware. The enforcement of these page transitions and memory ownership policies significantly enhances SEV's security guarantees:

\begin{itemize}
    \item \textit{Confidentiality:} Prevents data leakage by enforcing strict page ownership and disallowing unauthorized aliasing or mapping of guest pages to hypervisor-visible regions.
    
    \item \textit{Integrity:} Hardware validation using RMP and PVALIDATE defends against memory remapping, page injection, or rollback attacks. Illegal transitions (e.g., from Hypervisor to Guest-Valid) are blocked to ensure strong memory isolation.
    
    \item \textit{Availability:} Enforces a consistent and verifiable page lifecycle. By rejecting illegal transitions and corrupted mappings, SEV-SNP reduces the risk of denial-of-service conditions caused by malicious or buggy hypervisors.
    
    \item \textit{Attestability:} Enhances remote attestation by including per-page metadata, memory layout, and feature enablement (e.g., SNP activation). This allows remote verifiers to confirm not just encryption but the structural integrity of the guest environment.
\end{itemize}

By introducing page state enforcement, Reverse Map Table tracking, and fine-grained validation, SEV-SNP transitions SEV from a memory encryption solution to a complete memory protection and isolation model. This marks a major step toward full hardware-rooted trusted execution environments.

\section{Experiments}
\label{sec:experiments}

\color{purple}

% \section*{Progress Summary:}

% \begin{table}[h]
%     \centering
%     \begin{tabular}{|l|c|}
%         \hline
%         \textbf{Task} & \textbf{Status} \\ 
%         \hline
%         All basic guest lifecycle and execution ABIs are modeled & \checkmark \\ 
%         \hline
%         All key data structures are implemented & \checkmark \\ 
%         \hline
%         Extending based AMD-SEV model for AMD-ES &  \checkmark \\ 
%         \hline
%         Extending based AMD-SEV model for AMD-SNP &  \checkmark \\ 
%         \hline
%         Unit test cases &  \checkmark \\ 
%         \hline
%         Full integration and integrated test case &  \checkmark \\ 
%         \hline
%         Prepare for symbolic execution & \checkmark \\ 
%         \hline
%         Generating Confidentiality \& Integrity properties & $\triangleright$ \\ 
%         \hline
%         Symbolic execution for security properties & $\triangleright$ \\ 
%         \hline
%         Cache/TLB modeling & \texttimes \\ 
%         \hline
%     \end{tabular}
% \end{table}

\color{black}

We have implemented the AMD SEV-base model with AMD SEV ES and AMD SEV SNP extensions as well as properties using \textit{Rosette}~\cite{Rosette}. It incorporates symbolic variables, assertions, and solver-aided functions to formally verify security guarantees. Specifically, Section~\ref{sec:exp-design} shows the AMD SEV model  derived from design and property abstraction of publicly available AMD SEV specifications and reference implementations~\cite{amdreview}. Section~\ref{subsec:ros-conf},~\ref{subsec:ros-int}, and ~\ref{subsec:ros-avial} describes the AMD concrete CIA properties for CIA notions introduced in Section~\ref{sec:model-prop}. Verification is performed using Rosette’s symbolic execution engine, which employs the \textit{Z3}~\cite{de2008z3} SMT solver to explore execution paths and validate constraints. The experiments were conducted on a docker container based  environment on Apple M2 CPU with 16 GB RAM. We evaluated 14 confidentiality properties, 7 integrity properties, and 3 availability properties to ensure CIA  coverage for VM-based trusted execution environments. The full implementation, including the SEV model and properties, is available at: \url{https://github.com/UFESL/TEE-Checking/tree/stable/sev}.

\subsection{Modeling of AMD SEV Architecture}
\label{sec:exp-design}

% \section{Formal Modeling of AMD SEV Architecture}
% \label{sec:formal-modeling}

The design of SEV introduces a complex interaction between hardware-managed encryption, guest VM states, and hypervisor-exposed interfaces. Modeling these interactions is essential to reasoning about confidentiality and integrity under adversarial control, particularly given SEV’s threat model that assumes an untrusted hypervisor. Our modeling approach is guided by design-based and property-based abstraction. Rather than attempting to simulate SEV’s low-level microarchitectural behavior in full, we abstract the system at the level of its formal interface and state transitions, focusing on components that affect security-critical properties. Specifically, we capture how guest contexts are initialized, how encryption keys are managed and revoked, and how memory ownership and isolation are enforced. This enables us to formally define and verify security properties such as noninterference and invariant preservation over symbolic executions of SEV’s state machine.

We begin this section by describing the high-level guest lifecycle in SEV, including secure creation, execution, migration, and teardown phases. We then identify the key symbolic data structures that represent SEV’s state, such as the guest context table (GCTX), ASID table, and Guest State Machine. These structures serve as the foundation of our model and capture the evolving state of SEV-managed VMs across their lifecycle. Next, we model the SEV API calls that drive state transitions. These ABI functions such as \texttt{LAUNCH\_START}, \texttt{ACTIVATE}, \texttt{SEND\_START}, and \texttt{DECOMMISSION} must be symbolically executed to evaluate their effects on system state and enforce policy constraints. Finally, we describe how we model these components with support for symbolic execution and how it enables security property verification. Table~\ref{tab:model} summarizes the Rosette-based implementations developed to model the AMD SEV architecture. The reported lines of code exclude debug statements and comments. This structured modeling effort allows us to reason about the correctness and security of SEV’s design under a realistic threat model and provides the foundation for formal proofs of confidentiality and integrity in the presence of privileged adversaries.

\begin{table}[htp]
\centering
\caption{Summary of Rosette implementations for modeling the AMD SEV architecture.}
\vspace{-0.1in}
\begin{tabular}{|l|c|c|c|}
\hline
\textbf{Description} & \textbf{Lines of Code}  \\
\hline
Modeling of stateful data structures & 655  \\
Modeling of ABI functions and state transitions   & 450 \\
Test cases for abstract refinement  & 140 \\
Utility functions and logic for symbolic execution  & 115 \\
\hline
\end{tabular}
\label{tab:model}
\end{table}

% AMD SEV defines a guest VM lifecycle that transitions through secure creation, execution, migration, and teardown phases. Each stage ensures that guest data remains encrypted and isolated from unauthorized access, particularly from privileged software such as the hypervisor.

% During \textit{guest creation}, the SEV firmware initializes memory encryption using a VM-specific key and assigns a unique Address Space Identifier (ASID) to the guest. This ASID is used to tag and isolate encrypted memory pages, serving a similar function to Intel TDX's HKID.

% In the \textit{guest execution} phase, encrypted memory pages are mapped to the VM, and access enforcement is managed via the Page Encryption Bit (PEB). This prevents unauthorized access to guest memory by marking pages as encrypted and associating them with the correct ASID.

% \textit{Snapshotting or migration} securely transfers a guest’s encrypted memory to another platform while preserving encryption and integrity guarantees.

% During \textit{guest teardown}, the guest's encryption keys are securely deleted, and memory pages are zeroed or invalidated to prevent post-deallocation data leakage.

% To ensure confidentiality, SEV combines memory encryption, ASID-based access control, and secure I/O paths. VM memory is encrypted using a unique key, and each page is tied to a single guest context through the ASID. Side-channel protections are implicitly supported by denying plaintext memory access to untrusted software, while secure I/O ensures that data transferred between the guest and external devices is also encrypted.

\subsubsection{ Modeling of Stateful Data Structures}

To formally verify the correctness and security of SEV, we model its architecture using a set of symbolic stateful data structures. Each structure captures key aspects of the guest lifecycle, page ownership, memory encryption, and policy enforcement.

\begin{enumerate}
    \item \textbf{Guest context table (GCTX):} 
    Stores metadata for each VM, including fields such as \texttt{STATE, HANDLE, ASID, POLICY, VEK, TEK, TIK, NONCE, MS, LD}. It governs the lifecycle state of the guest and is central to tracking guest status.

    \item \textbf{ASID table (ASIDT):} 
    Maps ASIDs to guest handles, ensuring correct association between encryption contexts and VM identifiers. This table enforces isolation by preventing multiple guests from sharing the same ASID.

    \item \textbf{Page encryption table (PEB):} 
    Tracks the encryption status of physical memory pages. Each entry maps a physical address to a flag indicating whether it is encrypted or not, thus enforcing confidentiality and preventing data leakage.

    \item \textbf{Guest policy table:} 
    Encodes constraints specified by the VM owner, such as disabling debugging (\texttt{NODBG}), restricting key sharing (\texttt{NOKS}), enforcing encrypted state (\texttt{ES}), and disabling outbound migration (\texttt{NOSEND}). Policies are enforced throughout the VM lifecycle.

    \item \textbf{Guest state machine (GSTATE):} 
    Models the legal state transitions of a VM using a set of symbolic states: \texttt{UNINIT}, \texttt{LUPDATE}, \texttt{LSECRET}, \texttt{RUNNING}, \texttt{SUPDATE}, \texttt{RUPDATE}, and \texttt{SENT}. Each transition is triggered by ABI calls.

    \item \textbf{Memory encryption key table (MEKT):} 
    Maintains the association between ASIDs and VM encryption keys (VEKs), enabling the encryption and decryption of memory pages on a per-VM basis.
\end{enumerate}

\subsubsection{Modeling ABI Functions and State Transitions}

SEV defines a set of Application Binary Interface (ABI) functions that drive state transitions across the guest lifecycle. These functions must be modeled symbolically to reason about security CIA properties under adversarial control. These ABIs can be categorized into four major flows.

\paragraph{Guest Lifecycle ABIs:}
\begin{itemize}
    \item \texttt{LAUNCH\_START}: Initializes guest context and allocates encryption metadata.
    \item \texttt{LAUNCH\_UPDATE\_DATA}: Encrypts guest memory with the assigned VEK.
    \item \texttt{LAUNCH\_MEASURE}: Computes a cryptographic measurement for attestation.
    \item \texttt{LAUNCH\_SECRET}: Injects sealed secrets into guest memory.
    \item \texttt{LAUNCH\_FINISH}: Finalizes and locks the guest launch process.
\end{itemize}

\paragraph{Execution ABIs:}
\begin{itemize}
    \item \texttt{ACTIVATE}: Activates the VM by assigning an ASID and enabling memory encryption.
    \item \texttt{DEACTIVATE}: Deactivates the VM and releases the ASID and keys.
\end{itemize}

\paragraph{Migration ABIs:}
\begin{itemize}
    \item \texttt{SEND\_START}, \texttt{SEND\_UPDATE\_DATA}: Prepare encrypted guest state for transfer.
    \item \texttt{RECEIVE\_START}, \texttt{RECEIVE\_UPDATE\_DATA}: Restore encrypted state on the destination platform.
\end{itemize}

\paragraph{Teardown ABIs:}
\begin{itemize}
    \item \texttt{DECOMMISSION}: Securely deletes encryption keys and invalidates guest context.
\end{itemize}

\paragraph{Memory and Cache Operations:}
\begin{itemize}
    \item \texttt{SEV\_ASID\_ALLOC}, \texttt{SEV\_ASID\_FREE}: Manage ASID allocation and release.
    \item \texttt{SEV\_ENCRYPT\_PAGE}, \texttt{SEV\_DECRYPT\_PAGE}: Perform explicit memory encryption and decryption.
    \item \texttt{DF\_FLUSH}: Flushes data caches to eliminate residual plaintext data.
\end{itemize}

\begin{lstlisting}[caption={Guest Context Table Management in Rosette}]
(define GCTX (make-hash))

(define/contract GCTX-contract
  (hash/c integer? (list/c symbol? integer? integer? integer? integer? integer? integer? integer?))
  GCTX)

(define (add-guest handle state asid policy vek tek tik nonce ms)
  (hash-set! GCTX handle (list state asid policy vek tek tik nonce ms)))

(define (get-guest handle)
  (hash-ref GCTX handle #f))

(define (update-guest-state handle new-state)
  (when (hash-has-key? GCTX handle)
    (define guest (hash-ref GCTX handle))
    (hash-set! GCTX handle (list new-state (cadr guest) (caddr guest) (cadddr guest)
                                 (list-ref guest 4) (list-ref guest 5) (list-ref guest 6)
                                 (list-ref guest 7) (list-ref guest 8))))))
\end{lstlisting}

\begin{lstlisting}[caption={Guest Lifecycle API Management in Rosette}]
(define (LAUNCH_START handle asid policy)
  (define current-state (get-guest-state handle))
  (when (equal? current-state 'UNINIT)
    (add-guest handle 'LUPDATE asid policy 0 0 0 0 0) ;; Initialize with zeroed encryption keys
    (SEV_ASID_ALLOC handle asid)
    (set-guest-state handle 'LUPDATE)))

(define (LAUNCH_UPDATE_DATA handle memory-pages)
  (define current-state (get-guest-state handle))
  (when (equal? current-state 'LUPDATE)
    (for-each SEV_ENCRYPT_PAGE memory-pages) ;; Encrypt all pages
    (set-guest-state handle 'LSECRET)))
\end{lstlisting}

% \subsection{Extension of the model for AMD SEV ES and AMD SEV SNP}

% The initial model was developed based on the basic AMD  SEV spec but then to make it up to date AMD SEV we integrated AMD ES and AMD SNP details here.

% \textcolor{red}{Add flow diagram on snp and es}

% \begin{itemize}
%     \item SEV-ES
%     \begin{itemize}
%         \item Modeling encryption of encrypting and protecting all CPU register contents when a VM stops running.
%         \end{itemize}
%     \item  SEV SNP
%     \begin{itemize}
%         \item Reverse map table for integrity protection.
%         \item Extra page states and page validation for integrity protection.
%     \end{itemize}
% \end{itemize}

% \begin{table*}[h]
%     \centering
%     \begin{tabular}{|l|l|}
%         \hline
%         \textbf{Component} & \textbf{Modification} \\ \hline
%         GCTX & Added \texttt{encrypted-registers} field. \\ \hline
%         GSTATE & Updated to track SEV-ES states (VMEXIT/VMRUN). \\ \hline
%         LAUNCH\_MEASURE & Now verifies \textbf{register encryption} in addition to memory encryption. \\ \hline
%         VMCB & Added to store encrypted register states separately. \\ \hline
%         GHCB & Introduced to allow controlled register sharing with hypervisor. \\ \hline
%         SEND\_UPDATE\_DATA / RECEIVE\_UPDATE\_DATA & Updated to include encrypted register migration. \\ \hline
%     \end{tabular}
%     \caption{Modifications to Components for SEV-ES (The newvest)}
%     \label{tab:sev_es_modifications}
% \end{table*}

\subsubsection{Extension of the Model for AMD SEV-ES and AMD SEV-SNP}

The initial symbolic model described in this work is based on the SEV-base specification, which provides memory encryption and isolation for virtual machines. However, to support more recent hardware and stronger security guarantees, AMD has introduced two important extensions: \textit{SEV-ES} and \textit{SEV SEV-SNP}. We extend our original model to incorporate the additional mechanisms introduced by these extensions, capturing their impact on state transitions, system protection boundaries, and adversarial constraints.

% Figure~\ref{fig:sev_extensions_flow} presents a high-level flow diagram of SEV-ES and SEV-SNP integration, showing how these extensions add register encryption, page validation, and integrity enforcement into the guest lifecycle.

% \begin{figure}[htp]
%     \centering
%     \includegraphics[width=0.6\columnwidth]{figures/sev_es_snp_flow.png}
%     \caption{Extended model flow incorporating SEV-ES and SEV-SNP 
%     .}
%     \label{fig:sev_extensions_flow}
% \end{figure}

\paragraph{Register State Encryption and Hypervisor Control in SEV-ES}

SEV-ES enhances the original SEV model by encrypting the entire CPU register state when the guest is not executing. This prevents the hypervisor from observing sensitive data during VMEXIT or while accessing the VM’s virtual machine control block (VMCB). To model SEV-ES behavior, we introduce the following modifications:

\begin{itemize}
    \item A new \texttt{encrypted-registers} field is added to the guest context table (GCTX) to indicate that the guest operates under SEV-ES protection.
    \item The guest state machine (GSTATE) is extended to model \texttt{VMEXIT} and \texttt{VMRUN} transitions, ensuring that encrypted state transitions are captured formally.
    \item The \texttt{LAUNCH\_MEASURE} ABI is updated to account for register encryption verification in addition to memory encryption.
    \item A new component, the virtual machine control block (VMCB), is introduced to store encrypted CPU register states.
    \item The guest hypervisor communication block (GHCB) is added to support controlled data sharing between guest and hypervisor, modeling a limited and policy-driven communication interface.
    \item The \texttt{SEND\_UPDATE\_DATA} and \texttt{RECEIVE\_UPDATE\_DATA} ABIs are updated to include encrypted register migration, ensuring that guest state migration under SEV-ES includes the full encrypted execution context.
\end{itemize}

These updates ensure that the adversary model is constrained by the inaccessibility of CPU registers during guest transitions, aligning with SEV-ES's goal of removing hypervisor visibility into runtime state.

\paragraph{Memory Integrity and Nested Paging in SEV-SNP}

SEV-SNP extends SEV-ES with additional protections for memory integrity, preventing unauthorized modification or aliasing of guest memory pages. This is achieved through a hardware managed reverse map table (RMP), enforced page validation, and additional page metadata. Key enhancements we make to extend our model to support SEV-SNP include:

\begin{itemize}
    \item The Reverse Map Table is formally represented to track the ownership and validation state of each memory page.
    \item New page states are introduced into the model, enabling symbolic tracking of validated, unvalidated, and aliased pages.
    \item ABI commands that manipulate guest memory—such as \texttt{LAUNCH\_UPDATE\_DATA} and \\ \texttt{SEND\_UPDATE\_DATA} are extended to enforce RMP updates and validation checks.
    \item The model also accounts for page-level integrity violations, such as remapping attacks, and verifies that these are detected and halted at the hardware level.
\end{itemize}

By incorporating SEV-ES and SEV-SNP, our model captures the latest security guarantees provided by AMD’s evolving trusted execution architecture. These extensions are essential for verifying stronger confidentiality and integrity properties, especially under an untrusted hypervisor with full system control.

\vspace{0.1in}

% \color{red}
% \section{NoC and TEE}

% SEV-TIO does not talk about LLC cache message sharing. It talks about secure access to I/O.

% \begin{figure}[tp]
% \centering
% % \vspace{-0.1in}
% \includegraphics[width=0.90\columnwidth]{./figures/threat_model.png}
% % \vspace{-0.2in}
% \caption{Secure and Non-secure regions, confidential VMs and generic VMs sharing NoC and LLC.}
% \label{fig:threat_mod}
% % \vspace{-0.2in}
% \end{figure}

\color{black}

\subsubsection{Symbolic Execution Strategy}

Each ABI is modeled as a symbolic transition function over the global machine state. Transitions update one or more symbolic tables (e.g., GCTX, MEKT, GSTATE) and may be subject to preconditions such as policy rules or prior state. Symbolic execution enables exhaustive exploration of possible adversarial interactions with SEV APIs, supporting formal verification of CIA properties under untrusted hypervisor control.

\subsection{Modeling of Confidentiality Properties}
\label{subsec:ros-conf}

We modeled 14 confidentiality properties (CP), denoted as $CP_1$ through $CP_{14}$, based on the formal notion of confidentiality described in Section~\ref{subsec:conf-model}. Table~\ref{tab:confidentiality} summarizes these properties. The second column provides a brief description of the property. The third column indicates which version of AMD SEV each confidentiality property applies to—SEV-base, SEV-ES, or SEV-SNP. If a property applies to a lower-level SEV version, it is inherently satisfied by all subsequent extensions built on top of it (e.g., SEV-ES and SEV-SNP build on SEV-base). All listed confidentiality properties were successfully verified using our Rosette-based symbolic execution framework. Next, we discuss the first two confidentiality properties in detail.

\begin{table}[htp]
    \centering
    \renewcommand{\arraystretch}{1.2}
    \caption{Confidentiality properties for AMD SEV and its extensions.}
    \vspace{-0.1in}
    \begin{tabular}{|c|p{11cm}|c|}
        \hline
        \textbf{ID} & \textbf{Confidentiality Property (CP)} & \textbf{Applies to} \\
        \hline
        CP$_1$   & Memory encrypted by one ASID must not be decrypted using another ASID. & SEV-base \\
        CP$_2$   & No ASID reuse without explicit deallocation. & SEV-base \\
        CP$_3$   & Only the ASID owner may retrieve its VEK. & SEV-base \\
        CP$_4$   & Page access requires prior validation (via PVALIDATE). & SEV-SNP \\
        CP$_5$   & Only the assigned guest may access a page enforced by RMP checks. & SEV-SNP \\
        CP$_6$   & Register state must be encrypted on VMEXIT to prevent information leakage. & SEV-ES \\
        CP$_7$   & Validated pages must not be remapped to another guest. & SEV-SNP \\
        CP$_8$   & Guest must not use or modify page content unless it has been explicitly validated. & SEV-SNP \\
        CP$_9$   & Secrets must not be accessible after DEACTIVATE. & SEV-base \\
        CP$_{10}$  & Pages must not be reassigned without explicit unassignment. & SEV-SNP \\
        CP$_{11}$  & Secrets must only be injected after attestation is completed and the VM configuration is verified. & SEV-base \\
        CP$_{12}$  & Guest must not begin execution (RUNNING) without secrets and attestation. & SEV-SNP \\
        CP$_{13}$  & Secrets must not be visible after DECOMMISSION. & SEV-base \\
        CP$_{14}$  & GHCB must not leak secrets or sensitive guest state to the hypervisor. & SEV-ES \\
        \hline
    \end{tabular}
    \label{tab:confidentiality}
\end{table}

The first confidentiality property (CP$_1$) covers ASID memory encryption enforcement. Specifically, once a guest is assigned an ASID, all its memory pages must be encrypted and accessible only with the corresponding ASID and VEK. This property ensures that the memory pages remain confidential and inaccessible to unauthorized ASIDs. Equation~1 formally captures this property. Figure~\ref{lst:guest-access} shows Rossette implementation of the property for symbolic execution.

\vspace{-0.1in}

\begin{equation} \forall \text{guest}_i,\ \forall \text{page}_j \in \text{Pages}(\text{guest}_i),\ \forall \text{ASID}_k \neq \text{ASID}(\text{guest}_i),\ \neg \text{decrypt}(\text{page}_j, \text{VEK}_{\text{ASID}(\text{guest}_i)})
\end{equation}

The second confidentiality property (CP$_2$) covers no unauthorized ASID reuse. Once an ASID is assigned, it cannot be reused by another guest without being explicitly deallocated. Equation~2 formally captures this property.

% \begin{multline}
% \forall \text{guest}_i, \text{ if } \text{ASID}(\text{guest}_i) = A, \text{ then } 
% \forall t_1 < t_2, \\
% \neg \text{ASSIGN}(\text{guest}_j, A, t_2) \Rightarrow  \text{DEACTIVATE}(\text{guest}_i, A, t_1), 
%  \quad \text{for } i \neq j.
% \end{multline}

\begin{multline}
\text{CP}_2:\ \forall \text{guest}_i,\ \forall \text{guest}_j \neq \text{guest}_i,\ \forall t_1 < t_2, \\
\text{ASSIGN}(\text{guest}_j, A_2, t_2) \Rightarrow \exists t_1 < t_2\ \text{such that } \text{DEACTIVATE}(\text{guest}_i, A_1, t_1)
\end{multline}

If guest $j$ is assigned ASID ``$A_2$'' at time $t_2$, then there exists a time $t_1 < t_2$ such that guest$_i$ deactivated ASID ``$A_1$''. This prevents unauthorized ASID reassignment, which could lead to memory compromise. 

\begin{lstlisting}[caption={Rossette implementation of CP1: only the assigned guest may access the page}, label={lst:guest-access}]
(define (check-page-ownership-property)
    (define-symbolic attacker-id integer?)
    (define-symbolic page-addr integer?)
    
    (define legitimate-owner 42)
    (assign-page-to-guest page-addr legitimate-owner)
    (define attacker-constraint (not (= attacker-id legitimate-owner)))
    (define no-access-property (not (owns-page? attacker-id page-addr)))
    (define result (solve (assert (and attacker-constraint (not no-access-property)))))
    
    (if (unsat? result)
        (printf "PASS: No unauthorized guest can access")
        (begin
        (printf "FAIL: Unauthorized access IS possible!\n")
        (printf "Counterexample: attacker-id = ~a, page-addr = ~a"
                (evaluate attacker-id result)
                (evaluate page-addr result))))
\end{lstlisting}

\begin{table}[htp]
    \centering
    \renewcommand{\arraystretch}{1.2}
    \caption{Integrity properties for AMD SEV and its extensions.}
    \vspace{-0.1in}
    \begin{tabular}{|c|p{11cm}|c|}
        \hline
        \textbf{ID} & \textbf{Integrity Property (IP)} & \textbf{Applies To} \\
        \hline
        IP$_1$   & Guest must not skip mandatory launch states. & SEV-base \\
        IP$_2$   & Secrets must not persist through migration. & SEV-base \\
        IP$_3$   & State transitions must be atomic. & SEV-base \\
        IP$_4$   & PVALIDATE must fail if encryption or RMP ownership is incorrect. & SEV-SNP \\
        IP$_5$   & Guest must not rollback to a prior lifecycle state. & SEV-SNP \\
        IP$_6$   & VEKs must not be reassigned or mutated once issued. & SEV-base \\
        IP$_7$   & Page version must be monotonic to prevent rollback attacks. & SEV-SNP \\
        \hline
    \end{tabular}
    \label{tab:integrity}
\end{table}

\subsection{Modeling of Integrity Properties}
\label{subsec:ros-int}

We modeled 7 integrity properties, denoted as $IP_1$ through $IP_{7}$, based on the formal notion of integrity described in Section~\ref{subsec:int-model}. Table~\ref{tab:integrity} summarizes these properties. The second column provides a brief description of the property. The third column indicates which version of AMD SEV each confidentiality property applies to—SEV-base, SEV-ES, or SEV-SNP. If a property applies to a lower-level SEV version, it is inherently satisfied by all subsequent extensions built on top of it (e.g., SEV-ES and SEV-SNP build on SEV-base). All listed integrity properties were successfully verified using our Rosette-based symbolic execution framework. Next, we discuss the first two integrity properties in detail.

The first integrity property (IP$_1$) represents the property that guest state transitions must be valid. In other words, a guest must follow the correct SEV lifecycle states (UNINIT $\rightarrow$ LUPDATE $\rightarrow$ LSECRET $\rightarrow$ RUNNING) and must not skip any stage. Equation~\ref{eq:ip1} formally captures this constraint, where $S_1$, $S_2$,, $S_3$,, $S_4$ are UNINIT, LUPDATE, LSECRET, and RUNNING, respectively. This prevents privilege escalation attacks where a guest bypasses security checks. Figure~\ref{lst:guest-flow} shows Rossette implementation of the property for symbolic execution.

\begin{multline}
\label{eq:ip1}
\forall \text{guest}_i, \text{ if } \text{STATE}_i = S_1 \text{ at } t_1, \text{ then } \forall t_2 > t_1, 
\text{STATE}_i \notin \{S_3, S_4\} \text{ unless } S_2 \text{ has occurred.}
\end{multline}

% \textbf{Where:}
% \begin{align*}
% S_1 &= \text{UNINIT} \\
% S_2 &= \text{LUPDATE} \\
% S_3 &= \text{LSECRET} \\
% S_4 &= \text{RUNNING}
% \end{align*}

\begin{lstlisting}[caption={Rossette implementation of IP1: guest must not skip mandatory launch states.}, label={lst:guest-flow}]
(define (check-launch-state-integrity)
    (define-type State (enum UNINIT LUPDATE LSECRET RUNNING OTHER))
    (define max-states 4) 
    (define-symbolic guest-trace (list-of State) #:length max-states)
    
    (define (valid-lifecycle? trace)
      (match trace
        [(list 'UNINIT 'LUPDATE 'LSECRET 'RUNNING) #t][_ #f]))
    (define reached-running? (member 'RUNNING guest-trace))
    (define invalid-transition? (and reached-running? (not (valid-lifecycle? guest-trace))))
    (define result (solve (assert invalid-transition?)))
    
    (if (unsat? result)
        (printf " PASS: All reachable RUNNING states follow valid transitions")
        (begin
          (printf "FAIL: Guest reached RUNNING through invalid lifecycle!")
          (printf "Counterexample: guest trace = ~a" (evaluate guest-trace result))))
\end{lstlisting}

The second integrity property (IP$_2$) ensures confidentiality by enforcing that secrets injected into a guest on the source platform must not persist after migration. Specifically, any secrets provisioned during the launch process (e.g., via LAUNCH\_SECRET) must be purged when the guest context is transferred to a destination platform during live migration. This prevents leakage of sensitive data beyond its intended lifecycle and platform boundary. Equation~\ref{eq:ip3} formally captures this constraint.

\begin{equation}
\label{eq:ip3}
\forall \text{guest}_i,\ \forall t_1 < t_2,\ \text{if } \text{MIGRATE}(\text{guest}_i, t_1, t_2),\ \text{then } \forall s \in \text{Secrets},\ s \notin \text{State}(\text{guest}_i, t_2)
\end{equation}

% The second confidentiality property (IP$_2$) covers register encryption on VMEXIT (SEV-ES). When a guest VM exits (VMEXIT), its CPU registers must be encrypted before control is returned to the hypervisor. This prevents hypervisor-based register snooping by enforcing register encryption.

% \begin{multline}
% \forall \text{guest}_i, \text{ if } \text{STATE}_i = \text{VMEXIT}, \\
% \text{then } \text{encrypted-registers}_i = \text{True}.
% \end{multline}

% \subsubsection{iP$_3$: Attestation Must Be Completed Before Secret Injection}
% \textbf{Property:} Secrets should only be injected into a guest (LAUNCH\_SECRET) if its measurement (LAUNCH\_MEASURE) has been verified.

% \begin{multline}
% \forall \text{guest}_i, \text{ if } \neg \text{LAUNCH\_MEASURE}(\text{guest}_i), \\
% \text{then } \neg \text{LAUNCH\_SECRET}(\text{guest}_i).
% \end{multline}

% \textbf{Ensures:} Prevents injection of secrets into an unverified guest, mitigating rollback/replay attacks.

% \begin{table}[htp]
%     \centering
%     \renewcommand{\arraystretch}{1.2}
%     \caption{Availability properties for AMD SEV and its extensions.}
%     \vspace{-0.1in}
%     \begin{tabular}{|c|p{10.5cm}|c|}
%         \hline
%         \textbf{ID} & \textbf{Availability Property (AP)} & \textbf{Pass/Fail} \\
%         \hline
%         AP$_1$   & Hypervisor must regain control. & \cmark \\
%         AP$_2$   & Guest must eventually progress. & \xmark \\
%         \hline
%     \end{tabular}
%     \label{tab:availability}
% \end{table}

\begin{table}[htp]
    \centering
    \renewcommand{\arraystretch}{1.2}
    \caption{Availability properties for AMD SEV and its extensions.}
    \vspace{-0.1in}
    \begin{tabular}{|c|p{8.5cm}|c|c|}
        \hline
        \textbf{ID} & \textbf{Availability Property (AP)} & \textbf{Applies To} & \textbf{Pass/Fail} \\
        \hline
        AP$_1$   & Hypervisor must regain control. & SEV-base & \cmark \\
        AP$_2$   & Guest must eventually progress. & SEV-SNP & \xmark \\
        \hline
    \end{tabular}
    \label{tab:availability}
\end{table}

\subsection{Modeling of Availability Properties}
\label{subsec:ros-avial}

We modeled 2 availability properties, denoted as $AP_1$ and $AP_{2}$, based on the formal notion of availability described in Section~\ref{subsesc:avil-model}. Table~\ref{tab:confidentiality} summarizes these properties. The second column provides a brief description of the property. The fourth column indicates whether the property is satisfied (\textit{pass} or \textit{fail}) by the AMD SEV model outlined in Section~\ref{sec:exp-design}. Next, we discuss the two availability properties in detail.

% The first availability property (AP$_1$) covers \textit{host-side liveness}. In other words, it says the guest must not prevent the hypervisor from regaining control. 

% \begin{equation}
% \label{eq:ap1}
% \forall m_i.\; curr(m_i) = v \Rightarrow \exists j \ge i.\; curr(m_j) = h
% \end{equation}

% The second availability property (AP$_2$) covers \textit{guest-side liveness}. In other words, if the VM is in a ready state and not blocked internally, it must eventually reach a new state.

% \begin{equation}
% \label{eq:ap2}
% \forall m_i.\; curr(m_i) = h \land Ready(Sv(m_i)) \Rightarrow \exists j > i.\; curr(m_j) = v \land Sv(m_j) \ne Sv(m_i)
% \end{equation}

The first availability property (AP$_1$) ensures host-side liveness. Specifically, it guarantees that control eventually returns to the hypervisor. In other words, if the current machine state is executing in the guest context, the system must eventually transition back to the hypervisor. Equation~5 formally captures this property. Figure~\ref{lst:ap1} shows Rossette implementation of the property for symbolic execution.

\begin{equation}
\label{eq:ap1}
\forall m_i.; \text{curr}(m_i) = v \Rightarrow \exists j \ge i.; \text{curr}(m_j) = h
\end{equation}

\begin{lstlisting}[caption={Rossette implementation of AP1: host must eventually regain control.}, label={lst:ap1}]
(define (check-hypervisor-regains-control trace)
    (define violation
        (ormap
         (lambda (i)
           (define mi (list-ref trace i))
           (if (equal? (current-context mi) 'v)
               (not (ormap (lambda (j)
                             (equal? (current-context (list-ref trace j)) 'h))
                           (range i (length trace))))
               #f))
         (range (length trace))))

      (define result (solve (assert (not violation))))
    
      (if (unsat? result)
          (printf "PASS: Host regains control after each VM execution segment.")
          (begin
            (printf " FAIL: VM executes but hypervisor never regains control")
            (printf "Counterexample trace:")
           (for ([i (range (length trace))]) (printf "Step ~a: context = ~a\n" i (evaluate (current-context (list-ref trace i)) result))))))
\end{lstlisting}

The second availability property (AP$_2$) enforces guest-side liveness. If the hypervisor is currently executing and the guest is in a ready state (i.e., not internally blocked), then the guest must eventually make progress to a new state.

\begin{equation}
\label{eq:ap2}
\forall m_i.; \text{curr}(m_i) = h \land \text{Ready}(\text{Sv}(m_i)) \Rightarrow \exists j > i.; \text{curr}(m_j) = v \land \text{Sv}(m_j) \ne \text{Sv}(m_i)
\end{equation}

% \vspace{1em}

% \noindent \textbf{AP\textsubscript{3}: \textit{Termination Progress Guarantee}}\\
% \textbf{Property:} If a guest can terminate in a non-adversarial run, it must also eventually terminate under adversarial scheduling.
% \[
% Terminated(Sv(V_1)) \Rightarrow \lozenge Terminated(Sv(V_2))
% \]

As you can see only $AP_1$ is satisfied and $AP_2$ is not satisfied. Here $AP_1$ is refered to \textit{host-side liveness}. All SEV technologies mention of availability and guarantee that the hypervisor can regain control when it desires (e.g., via a physical timer interrupt) or terminate a guest at any time without the consent of that VM. The second notion of availability (\textit{guest-side liveness}) is whether the guest enjoys any guarantees of availability such as a minimum run-time and no-starvation. This is not part of any of the current SEV technology availability guarantees as a malicious hypervisor can choose not to run some or all of a guest VMs. 
%But a hardware-enforced trsuted scheduling, watchdog timer and detecting guest liveness violation through remote attestation can be considered as potential solutions to provide more control over textit{guest-side liveness} in AMD SEV.

\subsection{Completeness of the AMD SEV Model and CIA Properties}

We argue that our formal model and the set of security properties (confidentiality, integrity, and availability) are complete with respect to the intended behavior of the AMD SEV guest lifecycle. We provide intuitive justification by outlining all observable and security-relevant behaviors captured in the model and showing how each is either constrained by the operational semantics or verified via properties.

\subsubsection{Model Completeness}

Our formal model captures all security-relevant operations across the SEV lifecycle, including guest creation, memory assignment, attestation, execution, and teardown. It includes:

\begin{itemize}
    \item \textbf{State transitions:} Modeled using symbolic execution over guest and platform state (\texttt{GCTX}, \texttt{GSTATE}, \texttt{PAGEINFO}, etc.).
    \item \textbf{Lifecycle coverage:} Transitions span \texttt{UNASSIGNED}, \texttt{ASSIGNED}, \texttt{CONFIGURED}, \texttt{LAUNCHED}, \texttt{RUNNABLE}, \texttt{DEACTIVATED}, and \texttt{DECOMMISSIONED} states.
    \item \textbf{Hypervisor interaction:} Explicit attacker model simulates adversarial control over scheduling and memory, bounded by architectural constraints.
    \item \textbf{Cryptographic protections:} VEK and ASID scoping, RMP ownership, and page validation are modeled to enforce access and visibility restrictions.
\end{itemize}

Every security-sensitive behavior (e.g., memory injection, secret visibility, resource reassignment, lifecycle rollback, etc.) is  modeled and symbolically evaluated to ensure that potential violations can be expressed and caught. To ensure correctness of the model, we performed iterative model development with abstract refinement using concrete test cases and properties. 

\subsubsection{Property Completeness}

We designed a set of 23 security properties that cover all aspects of confidentiality, integrity, and availability in the SEV lifecycle. These properties are defined to check the following classes of behaviors:

\begin{itemize}
    \item \textbf{Confidentiality (14 properties):} Unauthorized access to secrets, memory, or register state (e.g., CP$_1$–CP$_{14}$).
    \item \textbf{Integrity (7 properties):} Unauthorized modification of memory or lifecycle state, and improper rollback or reuse (e.g., IP$_1$–IP$_7$).
    \item \textbf{Availability (2 properties):} Hypervisor preemptibility and guest liveness under adversarial scheduling (e.g., AP$_1$–AP$_2$).
\end{itemize}

Each property is mapped to one or more expected behaviors derived from the SEV threat model. Together, they ensure that the guest behaves correctly across all permitted transitions, and that any adversarial deviation is detected. Therefore, the combined model and property set provide a complete specification and verification framework for SEV security. All behaviors defined in the lifecycle specification are either enforced by state transition rules or validated by properties, ensuring full coverage of the guest execution and its interaction with a potentially malicious host. 
Table~\ref{tab:sym-ex-res} shows results for property verification for different types of properties. Our results demonstrate that we can verify confidentiality, integrity, and availability properties in a reasonable time.

\begin{table}[t]
\centering
\caption{Rosette symbolic execution and property verification results by property type.}
\label{tab:sym-ex-res}
\begin{tabular}{|l|c|c|c|}
\hline
\textbf{Property Type} & \textbf{Properties} & \textbf{Lines of Code} & \textbf{Verification Time (s)} \\
\hline
Confidentiality & $CP_1$--$CP_{14}$ & 650 & 20.52 \\
Integrity       & $IP_1$--$IP_7$    & 450 & 12.45 \\
Availability    & $AP_1$--$AP_2$    & 120 & 3.67 \\
\hline
\end{tabular}
\end{table}

\section{Conclusion}
\label{sec:conclusion}

This paper presented a comprehensive framework for the formal verification of VM-based TEEs, addressing the critical need for robust security mechanisms in the face of evolving threats. We have developed a formalization of confidentiality, integrity, and availability for confidential virtual machines (VM), proposing a secure and verifiable model in the context of powerful adversaries. Our contributions, including the formalization of a confidential VM, the establishment of formal definitions for confidentiality, integrity, and availability within VM-based TEEs, and the development of a refinement-based methodology, underline the importance and effectiveness of formal verification in ensuring the security of VM-based trusted execution environments. Our experimental results demonstrate the applicability and resilience of our framework to analyze sophisticated attack scenarios, highlighting its potential to significantly enhance the security posture. By proving the confidentiality and integrity guarantees of the AMD SEV platform through machine-checked proofs, we not only validate our approach but also pave the way for future research in securing virtualized TEE environments. By offering a scalable and flexible solution that meets the dynamic nature of cloud services, VM-based TEEs stand at the forefront of defending against unauthorized access and ensuring the confidentiality and integrity of critical data. Our work contributes to the ongoing efforts in securing these environments, providing a foundation for further advancements in the field of trusted computing.

\section*{Acknowledgments}
This work was partially supported by the Semiconductor Research Corporation (SRC) grant 2022-HW-3128 (Task 3128.001).

% \import{sections/}{points.tex}
\bibliographystyle{IEEEtran}
\bibliography{IEEEabrv,bibliography.bib}
% \newpage
% \import{sections/appendix}{points.tex}

\end{document}